\documentclass[aps,prl,twocolumn,groupedaddress]{revtex4}
\usepackage[dvips]{graphics, color}
\usepackage{amsmath}
\usepackage{amssymb}
\usepackage{float}
\usepackage{bm,times}
\usepackage{graphicx}
\newcommand{\be}{\begin{equation}}
\newcommand{\ee}{\end{equation}}
\newcommand{\bea}{\begin{eqnarray}}
\newcommand{\eea}{\end{eqnarray}}

\newcommand{\degree}{\ensuremath{^\circ}}
\newcommand*\mystrut[1]{\vrule width0pt height0pt depth#1\relax}
\DeclareFontFamily{U}{euc}{}
\DeclareFontShape{U}{euc}{m}{n}{<-6>eurm5<6-8>eurm7<8->eurm10}{}%
\DeclareSymbolFont{AMSc}{U}{euc}{m}{n}
\DeclareMathSymbol{\umu}{\mathord}{AMSc}{"16}

\begin{document}

\title{Control of Material Damping in High-$Q$ Membrane Microresonators}

\author{P.-L.~Yu}
\author{T.~P.~Purdy}
\author{C.~A.~Regal}
\affiliation{JILA, University of Colorado and National Institute of Standards and Technology, and
Department of Physics, University of Colorado, Boulder, Colorado 80309, USA}
\date{\today}

\begin{abstract}
We study the mechanical quality factors of bilayer aluminum/silicon-nitride membranes.  By coating ultrahigh-$Q$ ${\rm Si_3N_4}$ membranes with a more lossy metal, we can precisely measure the effect of material loss on $Q$'s of tensioned resonator modes over a large range of frequencies. We develop a theoretical model that interprets our results and predicts the damping can be reduced significantly by patterning the metal film. Using such patterning, we fabricate Al-${\rm Si_3N_4}$ membranes with ultrahigh $Q$ at room temperature. Our work elucidates the role of material loss in the $Q$ of membrane resonators and informs the design of hybrid mechanical oscillators for optical-electrical-mechanical quantum interfaces.
\\~\\PACS number(s): 03.67.-a, 42.50.-p, 85.85.+j, 46.40.Ff
\end{abstract}

\maketitle

Silicon nitride membranes have recently emerged as promising resonators for applications ranging from precision sensing to realization of a mesoscopic quantum harmonic oscillator~\cite{WilsonRae2007,Marquardt2007a,Thompson2007}.~Because of their large tensile stress, ${\rm Si_3N_4}$ membranes can have MHz resonant frequencies with sub-Hz damping rates.~The resulting room-temperature $Q$-frequency products of above $10^{13}~\mbox{Hz}$ approach the performance of quartz oscillators~\cite{Michael2007}.~This ultrahigh $Q$ combined with a two-dimensional geometry is an ideal platform for control and detection of motion in a high-finesse Fabry-P\'erot cavity, and cooling ${\rm Si_3N_4}$ membranes in such cavities to their quantum-mechanical ground state is a near-term prospect.~However, the mechanisms that limit the realized $Q$-factors of these tensioned resonators are just beginning to be explored~\cite{WilsonRae2011,Wilson2009,Zwickl2007,Jockel2011}.

To date, studies have focused on pure dielectric ${\rm Si_3N_4}$  membranes, but a variety of proposed cavity mechanics experiments would be enabled by the addition of a metallic layer to ${\rm Si_3N_4}$ while maintaining high $Q$ \cite{Teufel2011b,Regal2011,Taylor2011,Rabl2009}.~Foremost, a metallic membrane section could form a capacitor plate that couples to a microwave LC resonator; in fact, pure metallic drums have recently been ground-state cooled using a combination of cryogenic and microwave cavity cooling ~\cite{Teufel2011b}.  With a hybrid dielectric/metallic membrane, one could couple mechanical motion simultaneously to optical light and a microwave electrical circuit in the quantum regime~\cite{Regal2011}.~Such a device could solve the difficult, yet crucial, problem of transferring quantum states between microwave and optical photons.~It could also enable enhanced detection of excitations in a room-temperature electrical circuit via photodetection~\cite{Taylor2011}.~Further, a magnetic metallic film could be used for magnetic coupling of spins to membrane motion~\cite{Rabl2009}.~However, the success of these applications will hinge on creating hybrid membranes with a sufficiently high quality factor at relevant temperatures.

\begin{figure} \begin{center}
\includegraphics[scale=0.7]{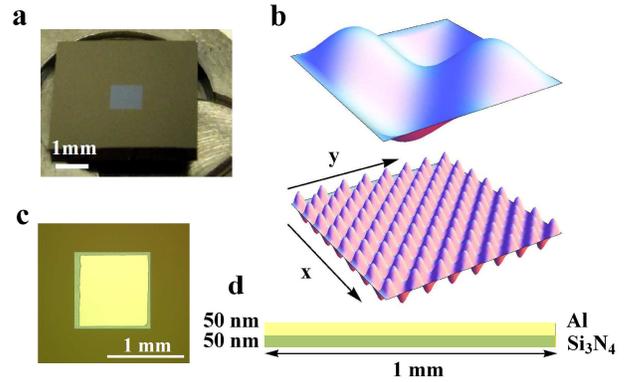} 
\caption{(color online). Geometry of membrane modes. (a) Image of a 1~mm membrane in its silicon frame. (b) Illustrations of the $(m,n)=(2,2)$ and $(15,15)$ modes. (c) Image of a patterned Al film on top of $\mathrm{Si_3N_4}$.  The central square is Al and the rim is $\mathrm{Si_3N_4}$ suspended on a Si frame. (d) Schematic diagram of the Al/$\mathrm{Si_3N_4}$ bilayer membrane (50 nm Al and 50 nm $\mathrm{Si_3N_4}$).} \label{fig1}
\end{center}
\end{figure}

In our work, we add metallic thin films to ${\rm Si_3N_4}$ membranes and explore the quality factor of many spatial modes of the membranes (Fig.~\ref{fig1}).  First, we identify two distinct loss mechanisms in our experiments: (1) loss of energy from the mechanical
mode into the substrate, i.e. radiation loss, and (2) material damping due to the lossy metallic film.  Then, we isolate the material loss-limited $Q$ and develop an anelastic theory that explains the observed dependence of $Q$ on frequency for a general clamped, lossy membrane.  Our work clarifies the role of material loss in highly-stressed two-dimensional resonators, and has significant predictive power.  Finally, we calculate and demonstrate that by removing the metal in a very small region near the clamp, we can create metallic ${\rm Si_3N_4}$ membranes with impressive quality factors of over $5\times10^6$ at 1~MHz at room temperature.

We use 50 nm-thick stoichiometric LPCVD nitride membranes that are supported by a 200~$\umu$m-thick silicon frame (from Norcada Inc.). The membranes are in a square geometry of side length $l\!\!=\!\!0.5$~mm or 1~mm with tensile stress $\sigma \!\!\sim\!\! 0.9$~GPa and mass density $\rho \!\!\sim\!\!2.7~\text{g}/\text{cm}^3$. The membrane mode shapes are given by approximately sinusoidal functions like those shown in Fig.~\ref{fig1}(b) with resonant frequencies $f_{mn}\!\sim\!\sqrt{\sigma(m^2+n^2)/4\rho l^2}$, where $m,n$ are the integer mode indices representing the number of antinodes.  The silicon frame is glued at three corners to a metal form on a piezoelectric actuator. To probe the mechanical displacement, we position the membrane at the end of one arm of a Michelson interferometer. We characterize the mechanical quality factor by monitoring the ringdown of the mechanical excitation as a function of time in vacuum of less than $10^{-6}$~torr.

\begin{figure} \begin{center}
\includegraphics[scale=0.63]{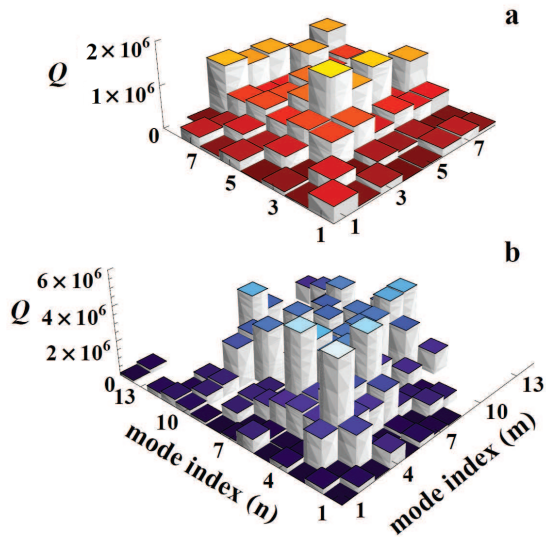} 
\caption{Radiation loss for a square membrane. Measurements of quality factors for many different modes of an (a) $0.5 \times 0.5$ mm and (b) $1 \times 1$ mm $\mathrm{Si_3N_4}$ membrane. The symmetric modes generally have higher $Q$ than asymmetric modes, as predicted by a radiation loss model \cite{WilsonRae2011}.} \label{fig2}
\end{center}
\end{figure}

\begin{figure} \begin{center}
\includegraphics[scale=0.6]{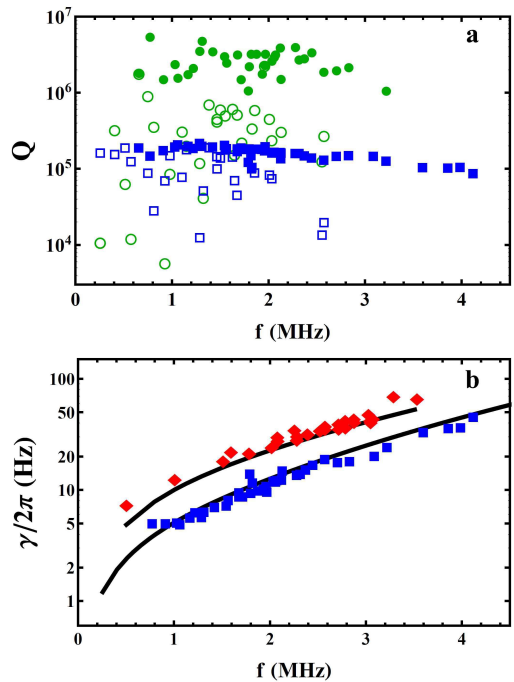} 
\caption{Extracting the material loss-limited $Q$. (a) Measured quality factors of a square  $\mathrm{Si_3N_4}$ membrane before (green circles) and after (blue squares) adding a 50 nm film of Al. The modes limited (not limited) by radiation loss are marked by open (closed) circles. The data are plotted as a function of frequency measured after adding the Al.  (b) Mechanical linewidth of the modes limited by the material loss of Al for $0.5 \times 0.5$ mm (red diamonds) and $1 \times 1$ mm (blue squares) membranes. To compare to theory, we calculate the damping rate $\gamma_{mn}/2\pi$ for each mode, and the points are connected with the displayed lines.} \label{fig3}
\end{center}
\end{figure}

In a first experiment, we measured the $Q$'s of pure $\mathrm{Si_3N_4}$ membranes. As shown in Fig.~\ref{fig2}, we have the ability to measure the quality factors of many modes (up to 150) with different symmetries and to confidently assign a mode $(m,n)$ to all measured points.   When the data are plotted versus resonant frequency (green circles in Fig.~\ref{fig3}), the $Q$ is non-monotonic.  However, when the data are plotted as a function of mode index in each dimension (Fig.~\ref{fig2}), we see that the asymmetric modes (indices $n$ and $m$ dissimilar) have strikingly smaller quality factors than do the more symmetric modes ($n$ and $m$ nearly equal).  This observation is consistent with expected trends for radiation loss of elastic waves through the membrane clamp. As recently calculated and measured in Ref.~\cite{WilsonRae2011}, the degree of destructive interference of elastic waves in the substrate is responsible for the symmetry dependence.  While we see consistently low $Q$'s with highly asymmetric modes, we see some variability due to the membrane mounting structure especially among the lowest order modes~\cite{Wilson2009}, as expected for a radiation loss mechanism~\cite{Jockel2011}.  However, for the high-order symmetric modes that asymptote to $Q$'s over a million, it becomes unclear whether radiation or material loss is the dominant effect.

In our next experiments, we deposit 50 nm of Al using e-beam evaporation on top of the pure $\mathrm{Si_3N_4}$ membrane measured in Fig.~\ref{fig2}(b). The membrane remains under large tensile stress, but adding the additional film does decrease the effective stress to $\sigma_{\rm{eff}}=0.35$~GPa.  With the addition of the metal, we see a drop in $Q$ to a maximum of $\sim\!2\times10^5$ as shown in Fig.~\ref{fig3}(a). Again, as a function of frequency, the $Q$'s are non-monotonic, but by drawing on our knowledge of the $Q$'s of the pure $\mathrm{Si_3N_4}$ membrane~[Fig.~2(b)], we can clearly distinguish radiation loss and material loss. The open squares in Fig.~\ref{fig3}(a) represent the asymmetric modes found to be radiation-loss limited for the pure $\mathrm{Si_3N_4}$ membrane. If we remove these points from the Al-$\mathrm{Si_3N_4}$ membrane measurements, we arrive at a clean set of points (closed squares) representing the material loss-limited $Q$ as a function of frequency.  Damping rates $\gamma=2\pi f/Q$ for two datasets obtained using this method are shown in Fig.~\ref{fig3}(b).

We have developed a theoretical framework to describe the frequency dependence of the material loss-limited quality factors of our two-dimensional structures.  We model the membrane as an anelastic plate that dissipates mechanical energy under cyclic loading~\cite{Zener1938}. Under oscillation, the material's strains and stresses are not in phase, and the energy supplied by the out-of-phase stresses is converted irreversibly to heat. This picture has been successfully developed to understand damping in one dimensional $\mathrm{Si_3N_4}$ strings \cite{Unterreithmeier2010, Schmid2011}.

In our case, we start by applying standard plate theory with an in-plane force~\cite{Leissa}, i.e.,~under tensile stress, to determine the normal modes. The modes must satisfy the boundary conditions of the clamped plate $W\!\!=\!\!(\partial/\partial x) W\!\!=\!\!0$ or $W\!\!=\!(\partial/\partial y) W\!\!=\!0$ for all four edges. We express the 2D mode function $W_{mn}(x,y)$ as a product of stressed-beam functions $u_m(x)u_n(y)$.  We have verified the accuracy of this description via perturbation theory~\cite{SM}.  We use a closed-form expression for the function $u_n(x)$ that is a sinusoid with an exponential correction near the edge for the clamped boundary condition.

For each mode, we can calculate the loss due to anelasticity. The oscillation of the plate induces oscillating strains $\varepsilon_{xx} e^{i\omega t}, \varepsilon_{yy} e^{i\omega t}$, and $\varepsilon_{xy} e^{i\omega t}$, and the accompanying stresses are given by  the usual constitutive equation of classical plate~\cite{Timoshenko} with the complex Young's modulus $\tilde{E}=E_1+iE_2$, where $E_2$ is called the loss modulus. During one cycle, the full expression for the energy lost is
\begin{equation}
\Delta U\!\!=\!\!\int \! \frac{2\pi E_2(x,y)}{1+\nu}\Big\{\frac{(\varepsilon_{xx}+\varepsilon_{yy})^2}{2(1-\nu)}+\frac{\varepsilon_{xy}^2}{4}-\varepsilon_{xx}\varepsilon_{yy}\Big\}dV
\end{equation}
where $\nu$ is the Poisson's ratio~\cite{SM}. Note, the strain term $\varepsilon_{xx}=-z(\partial^2 W/\partial x^2)$ is proportional to the curvature of the mode function. To calculate the quality factor, we also need an expression for the total stored energy. It can be obtained from the maximum kinetic energy~$U=2\rho\pi^2 f^2 \int {W(x,y)}^2dV$. The quality factor for a particular mode $W_{mn}$ is then given by $Q_{mn}=2\pi U_{mn}/\Delta U_{mn}$.

We start by using our theory to calculate the damping of fully-metallized membranes. We apply a least-squares fit to the two datasets (two different-sized membranes) in Fig.~\ref{fig3}(b) assuming a single frequency-independent loss modulus. This reveals an effective bilayer $E_2=0.55~\mathrm{GPa}$. The corresponding Al loss modulus is consistent with typical values for thin-film polycrystalline Al at room temperature, as measured, for example, via depositing Al on a low-loss Si cantilever \cite{Sosale2011}. The presumed microscopic origin of the loss is related to  crystallographic defects such as grain boundary sliding~\cite{Berry1981,Prieler1994} or kinks on dislocations~\cite{Hoehne2010}. Despite this underlying complexity, our model assumes very little about the microscopic origin of the loss.  Namely, we assume that the defects are uniformly distributed within the deposited metal in the $x$ and $y$ directions.  We also assume the temperature stays sufficiently constant in our measurements so as not to affect the loss modulus. We have verified that the heating due to our measurement laser of power 150 $\umu$W is not a significant effect by measuring constant quality factors as the power is varied from 10 to 900~$\umu$W.

With continued analysis of the theory we can not only model, but understand the $Q$ dependencies seen in Fig.~\ref{fig3}, and put our observations in the context of other studies in 1D and 2D \cite{WilsonRae2011,Jockel2011,Unterreithmeier2010,Schmid2011}.  We would like to understand: (1) The frequency dependencies, i.e.,~why an extremely corrugated mode has only a slightly lower $Q$ than the fundamental mode in our measurements (2) The geometry dependence, i.e.,~how damping should scale with resonator size.  First, we address the frequency dependence.  As noted above in the discussion of Eq.~(1), the loss is given by an integral of terms proportional to the mode curvatures.   We identify two contributions to the curvature, namely that induced at the clamped edge and that near the antinodes in the interior of the membrane.  If the curvature at the edge dominates we expect a flat $Q$ as a function of frequency, or if the antinode contribution dominates we expect a decreasing $Q$ as the frequency (and correspondingly the number of antinodes) increases.
We quantify these statements by deriving a simplified expression for $Q$ as a function of mode indices $m$ and $n$ for the limit of (1) an isotropic membrane, i.e. constant $E_2$ in $x$ and $y$ and (2) high-stress quantified by small  $\lambda m$ and $\lambda n$ where  $\lambda=\sqrt{E'h^2/3 \sigma l^2}$ is a dimensionless stress parameter.  Here $E'=E_1/(1-\nu^2)$ and $h$, $l$, and $\sigma$ are the height, length, and stress of the membrane respectively.  In these limits, Eq.~(1) becomes an integral over squared sinusoidal terms (antinode contribution) and an exponential term (edge contribution) to give a total $Q$ of~\cite{SM}
\begin{equation}
Q_{mn} \sim \frac{1}{\lambda}\frac{E_1}{E_2}
\bigg(\!
\underbrace{
\mystrut{1ex}1
}_{\mbox{edge}}\!\!
+
\underbrace{
\ \lambda\frac{(m^2+n^2)\pi^2}{4}
}_{\mbox{antinode}}
\bigg)^{-1}
\end{equation}
The term $\lambda (m^2+n^2)\pi^2/4$ determines whether there will be a frequency-dependent $Q$.  For our experiments, and similar experiments with large membranes~\cite{WilsonRae2011, Jockel2011}, $\lambda\sim10^{-4}\!\!-\!\!10^{-3}$, and hence we expect a relatively flat $Q$. However, if the edge length is decreased, $\lambda$ increases and the antinode contribution can become large.  Hence, a frequency dependence appears for experiments such as those in Ref.~\cite{Unterreithmeier2010} where shorter strings ($< 35$ $\umu$m) are used.

Further, the prefactor $1/\lambda$ in Eq.~(2) determines the geometry and
stress dependence for the $Q$ of the fundamental mode.  Physically,
$\lambda$ can be written as the ratio of bending energy to elongation energy~\cite{SM}, and as discussed in Ref.~\cite{Unterreithmeier2010},
exciting energy in the form of elongation energy rather than bending energy
leads to higher $Q$.  More concretely, based upon Eq.~(2), we predict that
if the membrane side length $l$ is doubled, $Q$ of the fundamental mode
will double for the same loss modulus, and this is exactly what is
observed in Fig.~\ref{fig3}(b); the analogous effect in 1D was observed in
Ref.~\cite{Unterreithmeier2010}.  We also see that as the stress is varied,
$Q$ scales with $\sqrt{\sigma}$, and hence the linewidth
$\gamma=2\pi f/Q$ remains constant.  While Eq.~(2) only holds in the stressed limit, a
calculation in the zero stress (flexural) limit reveals the linewidth
increases by a only a few factors from the highly-stressed case.

\begin{figure} \begin{center}
\includegraphics[scale=0.65]{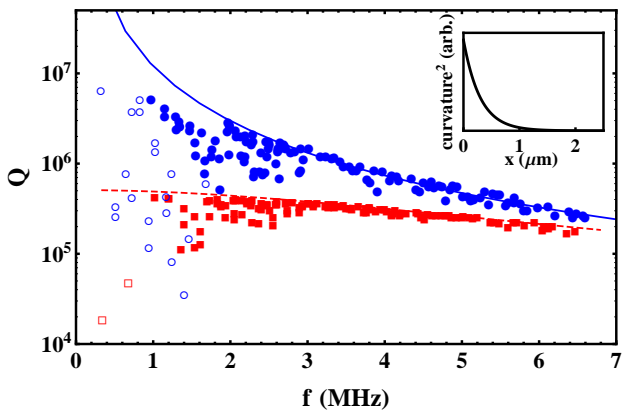} 
\caption{(color online). Ultrahigh-$Q$ metal-covered membranes. Measured quality factors of two $\mathrm{Si_3N_4}$ membranes with Al everywhere but near the edge of the membrane (blue circles)  [see Fig.~\ref{fig1}(c)] and a full film of Al as a control experiment (red squares). Asymmetric modes (with one mode index less than or equal to two) are marked by open circles or open squares. Calculated quality factors for each geometry are shown by the two lines; a single loss modulus is used for both.  (Inset)  The square of the curvature of a stressed mode as a function of distance along one coordinate of the membrane.  This function decays exponentially near the membrane edge.} \label{fig4}
\end{center}
\end{figure}

Our analysis above indicates that by making the loss modulus near the membrane edge small, we can reduce the loss significantly.  Using our ability to control the addition of material loss with the Al film, we can directly test this prediction. The inset to Fig.~\ref{fig1}(c) shows a $1 \times 1$~mm membrane where we deposited Al nearly everywhere except in a small $\sim$50~$\umu\mathrm{m}$ region near the edge.  The quality factors of this membrane were measured to be dramatically higher (blue circles) than a control experiment (red squares) in which an identical layer of Al was deposited everywhere on a separate membrane (Fig.~\ref{fig4}).  For these data we show the $Q$'s measured for all modes, but identify the lower-$Q$ asymmetric modes by open circles or squares.  For both datasets in Fig.~\ref{fig4}  we anneal the membranes at 340\degree C after depositing the Al film resulting in an effective stress of $\sigma_{\rm{eff}}=0.6$~GPa.  While annealing was not necessary for studying the fully-metallized membranes of Fig.~\ref{fig3}, the unequal stress of the Al film in the partially-metallized membrane makes the higher-order modes difficult to identify.  The annealing mitigates this problem, but we still cannot identify modes past 3~MHz. Hence past this point we analyze the $Q$ envelope by measuring the highest $Q$ mode in every 50~kHz window.

We can again apply our theory to quantitatively predict the $Q$ for this new geometry.  Since $E_2$ now becomes a function of position on the membrane, we return to using the full expression of Eq.~(1).  We assign a finite loss modulus for the Al region and zero loss for the $\mathrm{Si_3N_4}$ rim.  We approximate the mode functions as those expected for a uniformly stressed membrane.  This reveals the solid line in Fig.~\ref{fig4}, which we find scales as $1/n^2$ for diagonal modes, as expected from antinode contributions.  The dashed line shows the corresponding result for the fully metallized membrane using the same loss modulus.  (The loss modulus found here is  $E_2=0.3$ GPa, which is a smaller value than for Fig.~\ref{fig3} due to the annealing.)  We find the theory successfully traces out the envelope of the measured $Q$'s.  The lowest order modes of the partially-metallized membrane reach as high as $Q=6.5\times10^6$; this falls short of the predicted $Q$ just from Al material loss (blue line) likely because these modes are now again limited by radiation loss.  In comparison, recent measurements of metal microstrings at room temperature revealed $Q$'s of $10^3\!-\!10^5$~\cite{Pandey2010,Larsen2011}, and even at cryogenic temperatures, where the metal's material loss is significantly reduced, observed $Q$ values for tensioned microresonators are typically $10^5\!-\!10^6$ ~\cite{LaHaye2004,Regal2008,Teufel2011}.

It is elucidating to understand what would happen to the $Q$ trends we observe for the partially-metallized membrane upon varying the membrane stress. This requires analysis of the spatial dependence of the curvature in the membrane plane. In the inset to Fig.~\ref{fig4} we see the high-curvature area only occupies a very small $\sim\!\!1~\umu\mathrm{m}$ region near the edge of a high-tension membrane (here we use our lowest $\sigma_{\rm{eff}}=0.35$~GPa); specifically, the decay length is $\lambda l/4$~\cite{SM}.  As the stress is reduced (and hence $\lambda$ is larger), the curvature becomes more uniformly distributed over the membrane plane. Hence we would not expect a dramatic difference in $Q$ for a purely flexural mode when avoiding lossy material only at the edge.

The localized curvature of the tensioned membrane that we observe provides insight into a variety of membrane applications.  Note for higher-order two-dimensional modes the curvature varies along the edge of the membrane, i.e. there are low-curvature regions near the nodes at the membrane edge~\cite{SM}.  Thus, to create an electrical link between a central metallized patch and external circuits, and maintain high-$Q$ performance, one could tailor metal connections to match up with the low-curvature regions near nodes at the membrane edge~\cite{SM}. Further, membrane patterning via holes is a promising technique to increase reflectivity of membranes for optomechanics experiments, but like metal deposition, also has potential to introduce defects.  A full understanding of the curvature of the two-dimensional membrane is important for understanding the change in $Q$, or a lack of a decrease in $Q$, in recent patterning experiments~\cite{Kemiktarak2011,Bui2011}.

\acknowledgments{We thank I. Wilson-Rae, K. W. Lehnert, and R. W. Simmonds for valuable discussions and A. M. Kaufman for assistance. This work was supported by the DARPA QuASAR program, ONR YIP, and JILA NSF-PFC. CR thanks the Clare Boothe Luce Foundation for support. TP thanks the NRC for support.}

\onecolumngrid
\newpage

\setcounter{equation}{0}
\numberwithin{equation}{section}
\renewcommand{\theequation}{S\arabic{equation}}

\setcounter{figure}{0}
\numberwithin{figure}{section}
\renewcommand{\thefigure}{S\arabic{figure}}

\section{SUPPLEMENTARY INFORMATION}
\section{NORMAL MODES OF A HIGH-TENSION PLATE}

For an isotropic square plate with uniform boundary tension and under sinusoidal oscillation in the transverse direction, the dimensionless differential equation for a mode $W(x,y)$ is~[1]
\begin{align}
\label{plateEqn}
\frac{D}{\sigma h l^2}\biggr\{\frac{\partial^4 W}{\partial \xi^4}+\frac{\partial^4
W}{\partial \eta^4}\biggr\}-\left\{ \frac{\partial^2 W}{\partial
\xi^2}+\frac{\partial^2 W}{\partial \eta^2}\right\} +\underbrace{\frac{2D}{\sigma h l^2}\frac{\partial^4 W}{\partial \xi^2\partial
\eta^2}}_{\mathrm{perturbation \ term}}= \Omega^2 W
\end{align}
where $\xi=x/l$, $\eta=y/l$, $D$ is the flexural rigidity defined by $D=Eh^3/[12(1-\nu^2)]$, $E$ is the Young's modulus, $h$ is the plate thickness, $\nu$ is the Poisson's ratio, $\sigma$ is the inplane stress, $l$ is the length of the side of the plate, $\Omega=\omega\sqrt{\rho l^2/\sigma}$ is the dimensionless angular frequency, and $\rho$ is the mass per unit volume.

The modes for our device must satisfy the boundary conditions of a plate clamped on all four sides (CCCC), i.e.,
\begin{align}
\label{BCs}
W(\xi,\eta)\ \vline_{\mathrm{\ all\ edges}}=\frac{\partial  W(\xi,\eta)}{\partial n}\ \vline_{\mathrm{\ all\ edges}}=0
\end{align}
Here $n$ is the axis perpendicular to the edge.

To solve Eq.~(\ref{plateEqn}) and Eq.~(\ref{BCs}), note that the coefficient of the cross term is small for our devices: $2D/(\sigma h l^2)\sim 10^{-6}\ll 1$. Therefore, we can apply standard perturbation theory with $\epsilon\equiv2D/(\sigma h l^2)$ as the ``small parameter" and $\partial^4 /\partial \xi^2\partial \eta^2$ as the perturbation operator. Then the unperturbed equation can be solved by separation of variables.


Let $W(\xi,\eta)=X(\xi)Y(\eta)$, $\Omega^2=\Omega_{\xi}^2+\Omega_{\eta}^2$, and $E'=E/(1-\nu^2)$. We have
\begin{align}
&\frac{E' h^2}{12\sigma l^2}\frac{\partial^4 X}{\partial \xi^4}-\frac{\partial^2 X}{\partial \xi^2}=\Omega_{\xi}^2 X, \hspace{0.2 in} X\ \vline_{\ \xi=0,1}=\frac{\partial X}{\partial \xi}\ \vline_{\ \xi=0,1}=0 \notag\\
\label{eqny}
&\frac{E' h^2}{12\sigma l^2}\frac{\partial^4 Y}{\partial \eta^4}-\frac{\partial^2 Y}{\partial \eta^2}=\Omega_{\eta}^2Y, \hspace{0.25 in} Y\ \vline_{\ \eta=0,1}=\frac{\partial Y}{\partial \eta}\ \vline_{\ \eta=0,1}=0
\end{align}
These are the dimensionless equations of doubly clamped beams with Young's modulus $E'$and pre-stress $\sigma$ . The eigensolutions $u_m(\xi)$ (or $u_n(\eta)$) and the corresponding frequencies $\Omega_m$ (or $\Omega_n$) in the limit of small $\epsilon$ are given by
\begin{align}
&u_m(\xi)=
     \begin{cases}
       v_m(\xi),  & 0 \le \xi \le \frac{1}{2}\\
        (-1)^{m+1}v_m(1-\xi),  & \frac{1}{2} < \xi \le 1\\
     \end{cases}\label{umxi}\\
&v_m(\xi)=\sqrt{2a}\left\{\sin [\beta_m\xi]+\frac{\beta_m}{\alpha_m}\left(\exp[-\alpha_m\xi] -\cos [\beta_m\xi]\right)\right\}\label{vm}\\
&\Omega_{m}\sim m\pi\label{Om}
\end{align}
In Eq.~(\ref{vm}) $a\ll l$ is the vibration amplitude, and $ \alpha_m$ and $\beta_m$ are defined as
\begin{align}
 \alpha_m &
=\left(\frac{1}{\epsilon}\right)^{1/2}\left(1+\sqrt{1+2\epsilon\Omega_m^2}\right)^{1/2}\notag\\
\beta_m &
=\left(\frac{1}{\epsilon}\right)^{1/2}\left(-1+\sqrt{1+2\epsilon\Omega_m^2}\right)^{1/2}\label{alphabeta}
\end{align}
These functions have a sinusoidal shape with an exponential correction near the edge for the clamped boundary condition. Beam functions when $\epsilon$ is intermediate in size are given in Ref.~[2].

The products $\{u_m(\xi)u_n(\eta)\}$ are then the solutions of the unperturbed equation. The normal modes of the plate, however, must be symmetrized because of the square geometry of the plate
\begin{align}
  & W_{mn}^{(0)}(\xi,\eta) =
\begin{cases}
       \frac{1}{\sqrt{2}}\Big(u_m(\xi)u_n(\eta)+ u_n(\xi)u_m(\eta)\Big), & m>n\\
       \frac{1}{\sqrt{2}}\Big(u_m(\xi)u_n(\eta)- u_n(\xi)u_m(\eta)\Big), & m<n\\
        u_m(\xi)u_m(\eta), & m=n
\end{cases}\label{syMode}\\
&\Omega_{mn}^{(0)\ 2}=\Omega_m^2+\Omega_n^2
\end{align}

The orthogonality condition is
\begin{align}
\label{orthn}
&\frac{1}{a^2}\int^{1}_{0}\int^{1}_{0}W^{(0)}_{mn}(\xi,\eta)W^{(0)}_{m'n'}(\xi,\eta)d\xi d\eta=\delta_{mm'}\delta_{nn'}
\end{align}

\subsection{Perturbation Correction to First Order}
Now we will calculate the first order correction due to the perturbation term. Note that for $m\neq n$, $\Omega_{mn}^{(0)}=\Omega_{nm}^{(0)}$, i.e., the modes $W_{mn}^{(0)}(\xi,\eta)$ and $W_{nm}^{(0)}(\xi,\eta)$ are degenerate. Applying the degenerate perturbation theory, it turns out that the perturbation operator is diagonalized in the basis set $\{W^{(0)}_{mn}(\xi,\eta)\}$. Thus, we can easily calculate the eigenfrequencies to first order $\Omega_{mn}$ and the mode shapes with the first order correction $W_{mn}(\xi,\eta)$. Assuming $\beta_p/\alpha_p\ll 1$ for $p\in\{m,n,m',n'\}$, or equivalently $\epsilon\pi^2\mbox{max}\{m,m',n,n'\}^2/2\ll 1$, we have
\begin{align}
&\left|\frac{{\Omega_{mn}}^2-\Omega_{mn}^{(0)\ 2}}{\Omega_{mn}^{(0)\ 2}}\right|\sim\frac{\epsilon\pi^2m^2n^2}{m^2+n^2}\ll 1\notag\\
&\left|\frac{1}{a^2}\int^1_0\int^1_0 W_{mn}(\xi,\eta)W_{m'n'}^{(0)}(\xi,\eta)d\xi d\eta\right|\leq\frac{2\pi^2\epsilon^{3/2}mm'nn'}{\left|m^2+n^2-m'^2-n'^2\right|}\ll 1,&\Omega_{mn}^{(0)}\neq\Omega_{m'n'}^{(0)}
\end{align}
Therefore, we can use $\{W^{(0)}_{mn}(\xi,\eta)\}$ as the mode functions for our anelasticity calculation.

\subsection{Mode Functions of Bilayer Plate}
 For the fully or near fully metallized plates, we calculate mode functions by using a single-layer plate with thickness $100$~nm, an effective stress, Young's modulus, mass density, and Poisson's ratio. The effective Young's modulus $E=E_1=135$~GPa is the average of the Young's modulus of ${\rm Si_3N_4}$ and Al weighted by their thickness. The mass density $\rho$ of Al and ${\rm Si_3N_4}$ are very similar, and hence we use an effective mass density of $\rho \sim 2.7\ \mathrm{g/cm^3}$. We use an effective Poisson's ratio of $\nu=0.31$.

\section{ANELASTIC LOSS OF A HIGH-TENSION PLATE}
We place the $xy$ plane at middle plane of the plate and let the plate oscillate with a small amplitude at an angular frequency $\omega$ in the $z$-direction. The oscillation of plate $We^{i\omega t}$ induces a variation of local bending and overall elongation [3], and thus gives oscillating strains of
\begin{align}
\label{strainW}
\tilde{\pmb{\varepsilon}}(t)
\equiv
\left( {\begin{array}{c}
\tilde{\varepsilon}_{xx}(t) \\
\tilde{\varepsilon}_{yy}(t) \\
\tilde{\varepsilon}_{xy}(t) \end{array} } \right)
&=
\underbrace{
\left( {\begin{array}{c}
-z\frac{\partial^2}{\partial x^2}We^{i\omega t}\\
-z\frac{\partial^2 }{\partial y^2}We^{i\omega t}  \\
-2z\frac{\partial^2 }{\partial x \partial y}We^{i\omega t}
\end{array} } \right)
}_{\mbox{local bending}\sim\mathcal{O}(W)}
+
\underbrace{
\left( {
\begin{array}{c}
\frac{1}{2}\left(\frac{\partial}{\partial x}We^{i\omega t}\right)^2\\
\frac{1}{2}\left(\frac{\partial}{\partial y}We^{i\omega t}\right)^2  \\
0
\end{array} } \right)
}_{\mbox{elongation}\sim\mathcal{O}(W^2) }
\\
& \approx
\left( {\begin{array}{c}
-z\frac{\partial^2 W}{\partial x^2}\\
-z\frac{\partial^2 W}{\partial y^2} \\
-2z\frac{\partial^2 W}{\partial x \partial y}
\end{array} } \right)e^{i\omega t}
\equiv
\left( {\begin{array}{c}
\varepsilon_{xx} \\
\varepsilon_{yy} \\
\varepsilon_{xy} \end{array} } \right)
e^{i\omega t}
\equiv
\pmb{\varepsilon_0}e^{i\omega t}
\end{align}
The strain variation is mostly induced by pure bending, as shown in Eq.~(\ref{strainW}). The accompanying variation of stresses are given by  the usual constitutive equation of classical plate~[1] with the complex Young's modulus $\tilde{E}=E_1+iE_2$
\begin{align}
\tilde{\pmb{\sigma}}(t)
&\equiv
\left( {\begin{array}{c}
\tilde{\sigma}_{xx}(t)\\
\tilde{\sigma}_{yy}(t)\\
 \tilde{\tau_{xy}}(t) \end{array} } \right)
 =
\frac{\tilde{E}}{1-\nu^2}
\left({\begin{array}{ccc} 1 & \nu & 0 \\
                   \nu &  1 & 0 \\
                   0 & 0 & (1-\nu)/2 \end{array} }\right)
\tilde{\pmb{\varepsilon}}(t)
\equiv
\tilde{E}\mathbf{M}\tilde{\pmb{\varepsilon}}(t)
\end{align}

Note that the stresses can be separated into in-phase and out-of-phase terms as follows
\begin{align}
\Re[\tilde{\pmb{\sigma}}(t)]&=\Re[\tilde{E}\mathbf{M}\tilde{\pmb{\varepsilon}}(t)]=\Re[\tilde{E}\mathbf{M}\pmb{\varepsilon_0}e^{i\omega t}]\notag\\
&=\underbrace{E_1\mathbf{M}\pmb{\varepsilon_0}\cos (\omega t) }_{\mbox{in-phase}}  -\underbrace{E_2\mathbf{M}\pmb{\varepsilon_0}\sin (\omega t)}_{\mbox{out-of-phase}}
\end{align}
The mechanical work done per unit volume per oscillation is then given by:
\begin{align}
\Delta w
&=\oint \Re[\tilde{\pmb{\sigma}}^{\mathrm{T}}]\Re[\frac{d \tilde{\pmb{\varepsilon}}}{dt}]dt\nonumber\\
&=\int^{\frac{2\pi}{\omega}}_0  \left[E_1 \cos(\omega t)\pmb{\varepsilon_0}^{\mathrm{T}}\mathbf{M} -E_2 \sin(\omega t)\pmb{\varepsilon_0}^{\mathrm{T}}\mathbf{M}\right]\Re[\pmb{\varepsilon_0}\frac{d e^{i\omega t}}{dt}]dt\nonumber\\
\label{lossstrain}
&=\pi E_2\pmb{\varepsilon_0}^T\mathbf{M}\pmb{\varepsilon_0}\ge 0
\end{align}
This leads to a dissipation. Meanwhile, the bending energy supplied by the in-phase stresses is stored in the system.
\begin{align}
\label{storedE}
u_{\mbox{bending}}
&= \int^{\pi/2\omega}_0 \Re[\tilde{\pmb{\sigma}}(t)]\cdot \Re[\frac{d \tilde{\pmb{\varepsilon}}}{dt}]dt=E_1\pmb{\varepsilon_0}^T\mathbf{M}\pmb{\varepsilon_0}/2=\frac{E_1}{E_2}\frac{\Delta w}{2\pi}
\end{align}
Eq.~(\ref{storedE}) is consistent with the strain energy density of a pure bending plate derived in Ref.~[4]. We can see that the local dissipation $\Delta w$ is proportional to the density of stored bending energy. In other words, the bending energy supplied by the out-of-phase stresses is converted irreversibly to heat.

To find the total dissipation per cycle, we integrate Eq.~(\ref{lossstrain}) over entire volume of the plate and insert the strain-displacement relations [Eq.~(\ref{strainW})]. We have
\begin{align}
\label{bendE}
\Delta U &=\int\Delta w dV
=\int\pi E_2\pmb{\varepsilon_0}^T\mathbf{M}\pmb{\varepsilon_0}\ dV\notag \\
&=  \int\!\!\!\int\!\!\!\int \frac{\pi E_2(x,y)}{(1-\nu^2)} \Big\{(\varepsilon_{xx}+\varepsilon_{yy})^2-2(1-\nu)(\varepsilon_{xx}\varepsilon_{yy}-\varepsilon_{xy}^2/4)\Big\}dxdydz\\
&=\int \frac{2\pi E_2(x,y)}{(1+\nu)} \Big\{\frac{(\varepsilon_{xx}+\varepsilon_{yy})^2}{2(1-\nu)}+\frac{\varepsilon_{xy}^2}{4}-\varepsilon_{xx}\varepsilon_{yy}\Big\}dV\\
&=\int z^2dz\int\!\!\!\int \frac{\pi E_2(x,y)}{(1-\nu^2)}
\Bigg\{
\underbrace{\Bigg(\frac{\partial^2W}{\partial x^2} + \frac{\partial^2W}{\partial y^2}\Bigg)^2}_{ (\mbox{mean curvature})^2 }
 -2(1-\nu)\underbrace{\left[\frac{\partial^2W}{\partial x^2}\frac{\partial^2W}{\partial y^2}-\left(\frac{\partial^2W}{\partial x \partial y}\right)^2\right]}_{\mbox{Gaussian curvature} }
\Bigg\} dxdy\label{Bcur}
\end{align}
The total loss is given by the integration of curvatures over the plate plane.
\subsection{Stored Energy}
In order to evaluate the quality factor $Q=2\pi U/\Delta U$, we need to calculate the stored energy $U$ and energy loss $\Delta U$. We calculated $\Delta U$ above, and here we analyze the stored energy, which can be obtained from the maximum kinetic energy or the maximum elastic energy. The maximum kinetic energy is
\begin{align}
U_{\mathrm{kinetic}}=\frac{\rho h \omega^2 }{2}\int\!\!\!\int W(x,y)^2 dxdy=2\rho\pi^2 f^2 \int W(x,y)^2 dV\label{kinE}
\end{align}
The maximum elastic energy for a tensioned resonator is the sum of the displacement-induced elongation energy and the bending energy~[3].  These two energies are given by
\begin{align}
U_{\mathrm{elongation}}&=\frac{\sigma h}{2}\int\!\!\!\int\left\{\left({\frac{\partial W}{\partial x}}\right)^2+\left({\frac{\partial W}{\partial y}}\right)^2\right\}dxdy\\
U_{\mathrm{bending}}&=\int u_{\mathrm{bending}}dV=\int\frac{E_1}{E_2}\frac{\Delta w}{2\pi}dV
\end{align}

\subsection{Effective Loss Modulus}
Using the expressions for $U_{mn}$ and $\Delta U_{mn}$ above, and assuming $E_2(x,y)$ is constant over the metallized area, we can evaluate the quality factor or damping rate for each mode with only one unknown quantity $E_2$. In our data analysis we extract an effective $E_2$ for the $100~\mbox{nm}$ bilayer membrane using a least-squares fit to our data, as discussed in the main text. The loss modulus of Al can then be estimated by $E_2(h_{\mathrm{Si_3N_4}}+h_{\mathrm{Al}})/h_{\mathrm{Al}}$. Where $h_{\mathrm{Si_3N_4}}$ and $h_{\mathrm{Al}}$ are the thickness of the $\mathrm{Si_3N_4}$ and the Al, respectively.

\section{INTERPRETATION}

\subsection{Small \boldmath$\lambda m$ ($\lambda n$) Limit}
Define a dimensionless parameter $\lambda=\sqrt{2\epsilon}=\sqrt{E'h^2/3\sigma l^2}$. In the limit of small $\lambda m$, which holds for our membranes and other high-tension resonators~[3, 5-7], Eq.~(\ref{alphabeta}) becomes
\begin{align}
 \alpha_m &
=\left(\frac{1}{\epsilon}\right)^{1/2}\left(1+\sqrt{1+2\epsilon\Omega_m^2}\right)^{1/2}\sim\left(\frac{2}{\epsilon}\right)^{1/2}\equiv\frac{2}{\lambda}\notag\\
\beta_m &
=\left(\frac{1}{\epsilon}\right)^{1/2}\left(-1+\sqrt{1+2\epsilon\Omega_m^2}\right)^{1/2} \sim m\pi
\end{align}
The beam equations Eqs.~(\ref{umxi})-(\ref{Om}) become (back to coordinate $x$)
\begin{align}
u_m(x)&=
     \begin{cases}
       v_m(x),  & 0 \le x \le \frac{l}{2}\\
        (-1)^{m+1}v_m(l-x),  & \frac{l}{2} < x \le l\\
     \end{cases}\\
v_m(x)&=\sqrt{2a}\left\{\sin \Big[\frac{m \pi x}{l}\Big]+\frac{\lambda m\pi}{2}\left(\exp\Big[\frac{-x}{\lambda l/2}\Big] -\cos \Big[\frac{m \pi x}{l}\Big]\right)\right\}\label{vmx}\\
\omega_{m}&\sim \frac{m\pi}{l}\sqrt{\frac{\sigma}{\rho}}
\end{align}

\subsection{Isotropic Plates}
For a plate with uniform thickness and constant $E_2$, we can simplify Eq.~(\ref{Bcur}) using Green's theorem. The integration of Gaussian curvature is zero for a clamped plate (CCCC).
\begin{align}
&\int\!\!\!\int\left[\frac{\partial^2W}{\partial x^2}\frac{\partial^2W}{\partial y^2}-\left(\frac{\partial^2W}{\partial x \partial y}\right)^2 \right]dxdy=\oint\frac{\partial^2 W}{\partial y^2}\frac{\partial W}{\partial x}dy-\oint\frac{\partial^2W}{\partial x \partial y}\frac{\partial W}{\partial x}dx=0\notag
\end{align}
Thus  Eq.~(\ref{Bcur}) becomes,
\begin{align}
\label{curdiss}
\Delta U &=\frac{\pi E_2 h^3}{12(1-\nu^2)}\int\!\!\!\int \left(\frac{\partial^2W}{\partial x^2}+ \frac{\partial^2W}{\partial y^2}\right)^2 dxdy
\end{align}
Note $\partial^2 W/\partial x^2$ is the curvature of the mode in the $x$ direction, and $\partial^2 W/\partial y^2$ is the curvature in the $y$ direction. The sum of the two curvatures is the mean curvature.

\subsection{Curvature Induced near the Clamped Edge and around the Antinodes}
In our anelastic model, the local dissipation of a clamped plate is proportional to square of the mean curvature, as shown in Eq.~(\ref{curdiss}). Hence, it is important to understand its spatial dependence. Since our 2D normal modes are the product of 1D modes, we start by calculating the curvature of a 1D mode $u_n$. In the limit of $\lambda n\ll 1$, we can approximate the square of the curvature as
\begin{align}
\left(\frac{d^2 u_{n}(x)}{d x^2}\right)^2
&\sim\frac{2n^2\pi^2 a}{l^4}
\left(
\frac{4}{\lambda^2}\exp\Big[\frac{-x}{\lambda l/4}\Big] +
n^2\pi^2\sin^2\left[\frac{n\pi x}{l}\right]
\right)& 0\le x\le \frac{l}{2}
\end{align}
The equation above describes the $x$ dependence of the curvature in the $x$ direction. Note that the coefficient of the first term is much larger than that of the second term. This first exponential term is concentrated in a small range $\lambda l/4= h\sqrt{E'/48\sigma}$, leading to significant loss near the edge. The second term distributes over the entire length with sinusoidal dependence, just like the curvature of a simply-supported beam.

\begin{figure} \begin{center}
\includegraphics[scale=0.30]{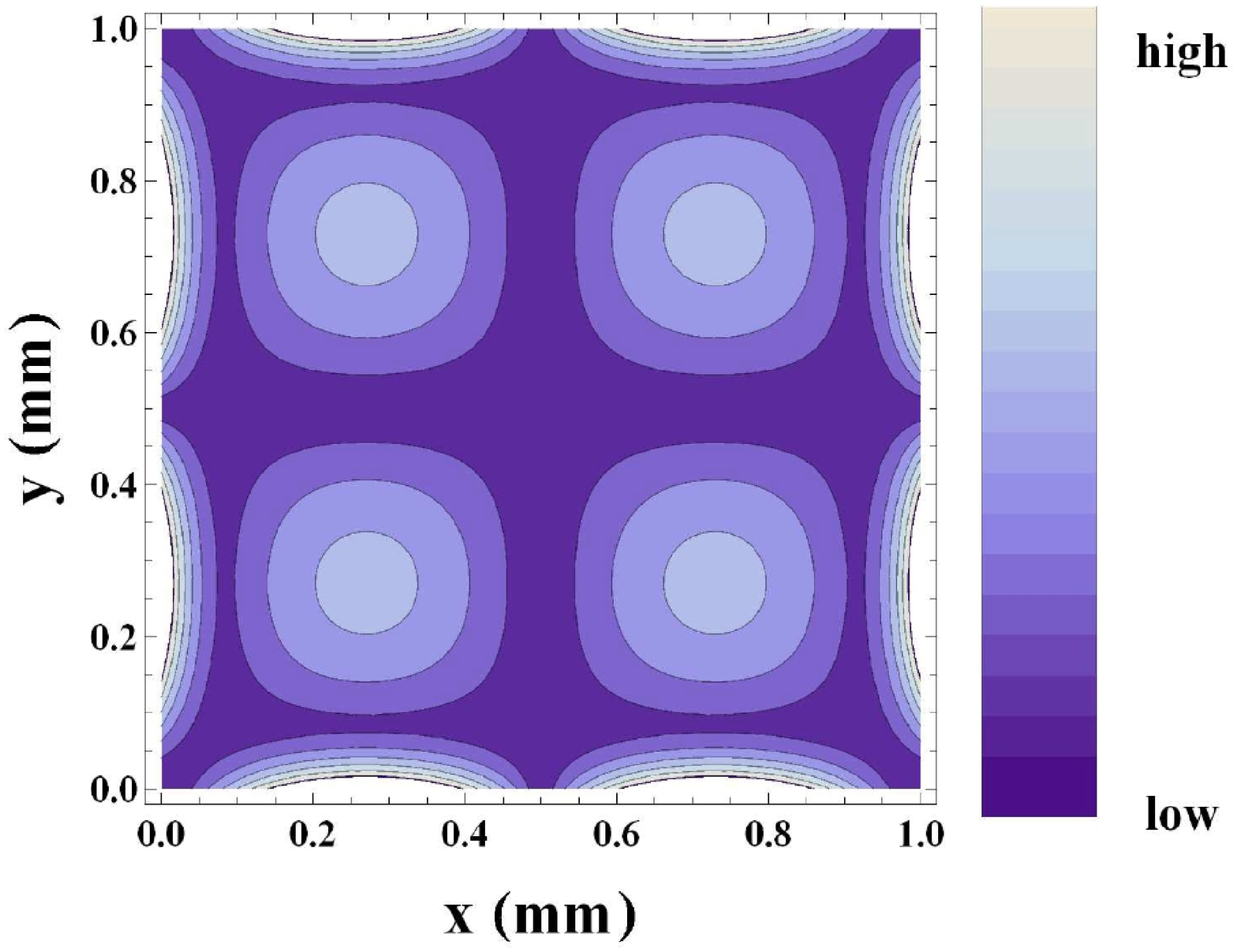} 
\caption{The spatial dependence of the mean curvature square of mode $(2,2)$ over entire plate plane. A low stress of 0.1~MPa is used and the values of curvature are scaled nonlinearly to make the structure more visible.} \label{fig5}
\end{center}
\end{figure}

For the 2D case, in addition to the above-mentioned features, the large curvature induced near the edge oscillates sinusoidally along the plate edge. To see that, we calculate the  mean curvature square of the diagonal mode $(n,n)$ at the edge $x=0$

\begin{align}
\label{2Dc}
\left(\frac{\partial^2 W_{nn}}{\partial x^2}+\frac{\partial^2 W_{nn}}{\partial y^2}\right)^2\ \vline_{\ x=0}
\sim \frac{16 n^2\pi ^2 a^2 }{\lambda ^2 l^4}\sin^2 \left[\frac{n\pi y}{l}\right]
\end{align}
See Fig.~\ref{fig5} for an illustration of square of the mean curvature over the entire plate plane.

\subsection{Edge Loss and the Frequency Dependence of the Quality Factor}
To understand the role of loss induced near the clamped edge, here we study $Q$ of a mode $(m,n)$ in the limit of small $\lambda n$ and $\lambda m$. First, we calculate the maximum kinetic energy. Applying Eq.~(\ref{orthn}) to Eq.~(\ref{kinE}), we have
\begin{align}
\label{keg}
U_{mn}\sim\rho h a^2l^2(\omega_m^2+\omega_n^2)/2\sim\sigma h a^2(m^2+n^2)\pi^2/2
\end{align}
 Second, we calculate the loss per cycle drawing on our earlier curvature analysis. Here we use $W_{mn}(x,y)=u_m(x)u_n(y)$. We have verified that the result is the same as using the symmetrized mode functions in Eq.~(\ref{syMode}).
\begin{align}
\Delta U_{mn}
&= \frac{\pi E_2}{1-\nu^2}\int z^2dz\int\!\!\!\int \left(\frac{\partial^2 W_{mn}}{\partial x^2}+\frac{\partial^2 W_{mn}}{\partial y^2}\right)^2 dxdy\notag\\
&= \frac{\pi E_2 h^3}{12(1-\nu^2)}\int\!\!\!\int \left(\frac{d^2 u_{m}(x)}{d x^2} u_n(y) +u_m(x)\frac{d^2u_{n}(y)}{d y^2}\right)^2 dxdy\notag\\
&= \frac{\pi E_2 h^3}{12(1-\nu^2)}
\left\{\int \left(\frac{d^2 u_{m}(x)}{d x^2}\right)^2 dx\int u_{n}(y)^2dy
+\int u_{m}(x)^2dx\int \left(\frac{d^2 u_{n}(y)}{d y^2}\right)^2 dy
\right.\notag\\ &\ \ \ \ \ \ \ \ \ \ \ \ \ \ \ \ \ \ \ \ \ \ \ \ \ \ \ \ \ \left.
+2\int \frac{d^2 u_{m}(x)}{d x^2}u_{m}(x) dx\int \frac{d^2 u_{n}(y)}{d y^2}u_{n}(y) dy\right\}\notag\\
&= \frac{\pi E_2 h^3}{12(1-\nu^2)}
\left\{
2 a l\int_0^{l/2} \left(\frac{d^2 u_{m}(x)}{d x^2}\right)^2 dx
+2 a l\int_0^{l/2} \left(\frac{d^2 u_{n}(y)}{d y^2}\right)^2 dy
\right.\notag\\ &\ \ \ \ \ \ \ \ \ \ \ \ \ \ \ \ \ \ \ \ \ \ \ \ \ \ \ \ \ \left.
+8 \int_0^{l/2} \frac{d^2 u_{m}(x)}{d x^2}u_{m}(x) dx\int_0^{l/2} \frac{d^2 u_{n}(y)}{d y^2}u_{n}(y) dy
\right\}\notag\\
&\sim
\frac{\pi E_2 h^3}{12(1-\nu^2)}
\bigg\{
\underbrace{
\frac{16m^2\pi^2a^2}{l^3\lambda^2}\int_0^{l/2} \exp\Big[\frac{-x}{\lambda l/4}\Big] dx
+\frac{16n^2\pi^2a^2}{l^3\lambda^2}\int_0^{l/2} \exp\Big[\frac{-y}{\lambda l/4}\Big] dy
}_{\mbox{``edge loss''}}
\bigg.\notag\\ &\ \ \ \ \ \ \ \ \ \ \ \ \ \ \ \ \ \ \ \ \ \ \ \ \ \ \ \ \ \bigg.
+\underbrace{
\frac{4m^4\pi^4a^2}{l^3}\int_0^{l/2} \sin^2\left[\frac{m\pi x}{l}\right] dx
+\frac{4n^4\pi^4a^2}{l^3}\int_0^{l/2} \sin^2\left[\frac{n\pi y}{l}\right] dy
}_{\mbox{``antinode loss''}}
\bigg.\notag\\ &\ \ \ \ \ \ \ \ \ \ \ \ \ \ \ \ \ \ \ \ \ \ \ \ \ \ \ \ \ \bigg.
+\underbrace{
8\Big(-\frac{2m\pi^2a}{l^2} \int_0^{l/2} \sin^2\left[\frac{m\pi x}{l}\right] dx\Big )
\Big(-\frac{2n\pi^2a}{l^2} \int_0^{l/2} \sin^2\left[\frac{n\pi y}{l}\right] dy\Big )
}_{\mbox{``antinode loss''}}
\bigg\}\notag\\
&\sim
\frac{\pi E_2 h^3}{12(1-\nu^2)}
\bigg\{
\underbrace{
\frac{4(m^2+n^2)\pi^2a^2}{l^2\lambda}
}_{\mbox{edge}}+
\underbrace{
\frac{(m^2+n^2)^2\pi^4a^2}{l^2}
}_{\mbox{antinode}}
\bigg\}\\
\label{analoss}
&\sim\frac{ (m^2+n^2)\pi^3E_2 h^3 a^2}{3(1-\nu^2)l^2\lambda}
\bigg\{\!
\underbrace{
\mystrut{1ex}1
}_{\mbox{edge}}
\!\!+
\underbrace{
\lambda\frac{(m^2+n^2)\pi^2}{4}
}_{\mbox{antinode}}
\bigg\}
\end{align}
We see that the ratio of the antinode loss to the edge loss is given by $\lambda (m^2+n^2)\pi^2/4$. Combining Eq.~(\ref{keg}) and (\ref{analoss}), we obtain
\begin{align}
\label{qfdep}
Q_{mn}=\frac{2\pi U_{mn}}{\Delta U_{mn}}\sim\frac{3(1-\nu^2)\sigma l^2\lambda}{E_2 h^2}\left(1+\lambda\frac{(m^2+n^2)\pi^2}{4}\right)^{-1}
\sim\frac{1}{\lambda}\frac{E_1}{E_2}\bigg(\!
\underbrace{
\mystrut{1ex}1
}_{\mbox{edge}}
\!\!+
\underbrace{
\lambda\frac{(m^2+n^2)\pi^2}{4}
}_{\mbox{antinode}}
\bigg)^{-1}
\end{align}
Similarly, we can derive the formula for a 1D string (with $\nu=0$)
\begin{align}
\label{qfdep1D}
Q_{n}\sim\frac{1}{\lambda}\frac{E_1}{E_2}\left(1+\lambda\frac{n^2\pi^2}{4}\right)^{-1}
\end{align}
For small $\lambda$ as low as $10^{-4}-10^{-3}$ in our devices and Ref.~[5-7], Eq.~(\ref{qfdep}) and Eq.~(\ref{qfdep1D}) correspond to a slightly decreasing $Q$ as a function of frequency. For $\lambda\sim2\times10^{-2}$ as in Ref.~[3], the edge loss and the antinode loss are comparable when $n=4$. This gives a stronger dependence on $n$, and hence frequency.

\subsection{Physical Meaning of $\lambda$}
To investigate the physical meaning of $\lambda$, we calculate the elongation energy of $(m,n)$ mode in the limit of small $\lambda n$:
\begin{align}
U_{\mathrm{elongation}}&=\frac{\sigma h}{2}\int\!\!\!\int\left\{\left(\frac{d u_{m}(x)}{d x}\right)^2 u_n(y)^2 +u_m(x)^2\left( \frac{d u_{n}(y)}{d y}\right)^2\right\} dxdy\notag\\
&=\sigma h l a\left\{\int^{l/2}_0 \left(\frac{d v_{m}(x)}{d x}\right)^2 dx
+\int^{l/2}_0 \left(\frac{d v_{n}(y)}{d y}\right)^2 dy\right\}\notag\\
&\sim\frac{2\pi^2\sigma h a^2}{l}\left\{m^2\int^{l/2}_0\cos^2\left[\frac{m \pi x}{l}\right]dx+n^2\int^{l/2}_0\cos^2\left[\frac{n \pi y}{l}\right]dy\right\}\notag\\
&=\pi^2\sigma h a^2(m^2+n^2)/2=U_{\mathrm{kinetic}}=U_{\mathrm{elastic}}
\end{align}
This means that the maximum elastic energy is dominated by the elongation energy. Thus,
\begin{align}
Q_{mn}=\frac{2\pi U_{mn}}{\Delta U_{mn}}\sim\frac{2\pi U_{\mathrm{elongation}}}{2\pi (E_2/E_1)U_{\mathrm{bending}}}=\frac{E_1}{E_2}\frac{U_{\mathrm{elongation}}}{U_{\mathrm{bending}}}
\end{align}
Comparing with Eq.~(\ref{qfdep}), we find
\begin{align}
\frac{U_{\mathrm{elongation}}}{U_{\mathrm{bending}}}&=\frac{1}{\lambda}\left(1+\lambda \frac{(m^2+n^2)\pi^2}{4}\right)^{-1}\notag\\
&\sim\frac{1}{\lambda}, \ \ \ \ \mbox{for}~\ (n,m)=(1,1)
\end{align}
Thus, $\lambda$ is the ratio of the bending energy to the elongation energy for the fundamental mode.

\section{REFERENCES}
[1] A.~W. Leissa, {\em Vibration of Plates} (NASA, Washington, D.C., 1969).

[2] A. Bokaian, J. Sound and Vib. {\bf 142},  481  (1990).

[3] Q.~P. Unterreithmeier, T. Faust, and J.~P. Kotthaus, Phys. Rev. Lett. {\bf
  105},  027205  (2010).

[4] S. Timoshenko, {\em Vibration Problems in Engineering} (D. Van Nostrand
  Company, Inc., New York, 1937).

[5] I. Wilson-Rae {\it et~al.}, Phys. Rev. Lett. {\bf 106},  047205  (2011).

[6] A. J\"{o}ckel {\it et~al.}, Appl. Phys. Lett. {\bf 99},  143109  (2011).

[7] S. Schmid, K.~D. Jensen, K.~H. Nielsen, and A. Boisen, Phys. Rev. B {\bf 84},
  165307  (2011).

\end{document}